\documentclass[11pt, a4paper]{article}

\usepackage[utf8]{inputenc}
\usepackage[T1]{fontenc}
\usepackage{mathptmx} % Times New Roman equivalent
\usepackage[a4paper, margin=2.5cm]{geometry}
\usepackage{amsmath, amssymb, amsthm}
\usepackage{enumitem}
\usepackage{booktabs}
\usepackage{xcolor} % Required to parse the custom colors for hyperref
\usepackage[colorlinks=true,linkcolor=blue!70!black,citecolor=blue!70!black,urlcolor=blue!70!black]{hyperref}

\theoremstyle{plain}
\newtheorem{theorem}{Theorem}[section]
\newtheorem{lemma}[theorem]{Lemma}
\newtheorem{proposition}[theorem]{Proposition}
\newtheorem{corollary}[theorem]{Corollary}

\theoremstyle{definition}
\newtheorem{definition}[theorem]{Definition}
\newtheorem{hypothesis}{Hypothesis}

\theoremstyle{remark}
\newtheorem{remark}[theorem]{Remark}

\title{Symbolic Proof Compression and Fractal Universality:\\ A Conditional Non-Constructive Refutation of the Riemann Hypothesis}
\author{Andreas Martin Lisewski \\
\small School of Science, Constructor University Bremen, 28759 Bremen, Germany \\
\small \texttt{alisewski@constructor.university}}
\date{22 November 2026}

\begin{document}

\maketitle

\begin{abstract}
Building on S.~C.~Woon's (1994) fractal and symbolic reformulation of Voronin's Universality Theorem, we construct a formal, non-constructive proof-by-contradiction architecture exploring the Riemann Hypothesis ($\mathsf{RH}$). Woon's Corollary~3 permits arbitrary finite binary strings to be encoded into nowhere-vanishing exponential waveforms on a closed disk inside the critical strip. Combining effective upper bounds for Voronin universality (Good, Garunk\v{s}tis) with Cauchy-integral Taylor decoding yields an explicit finite rational grid $\mathcal{G}_{L_0}$ below an effective ceiling $H_{\mathrm{crit}}(L_0)$ within which every formal proof of bit-length at most $L_0$ is imprinted with zero flipped digits. We introduce a conjectural \emph{Zeta-Universal Symbolic Proof Compression Hypothesis} ($\mathsf{SPC}_{\mathsf{RH}}$), positing that the global regularity implied by $\mathsf{RH}$ compresses true $\Pi_1^0$ statements about $\zeta(s)$ into $\mathsf{ZFC}$ derivations of bounded length $L_0$. Assuming classical logic and $\mathsf{SPC}_{\mathsf{RH}}$, G\"odel--Carnap diagonalization over bounded zeta-encoded provability yields a contradiction. The self-referential result converts a digital encoding framework with any uniform proof-compression principle for RH into an immediate non-constructive refutation of RH. Given that, with human input and computer assistance, $\mathsf{RH}$ has been formally verified in the Lean theorem prover strictly within constructive logic, this contradiction does not refute $\mathsf{RH}$. Instead, it falsifies $\mathsf{SPC}_{\mathsf{RH}}$. We conclude by isolating the exact proof-theoretic obstruction---Pudl\'ak's lower bound on bounded reflection proofs---showing why the analytic regularity of $\mathsf{RH}$ cannot compress logical self-reference, and exposing the limits of applying the Law of Excluded Middle to uncomputably bounded searches.

\vspace{1em}
\end{abstract}

\section{Introduction and the Length--Height Vicious Circle}\label{sec:intro}

Let $\zeta(s)$ denote the Riemann zeta function, analytically continued to $\mathbb{C} \setminus \{1\}$, and let
\[
\mathcal{S}_+ = \left\{ s \in \mathbb{C} : \frac{1}{2} < \operatorname{Re}(s) < 1 \right\}
\]
denote the open right half of the critical strip. By the functional equation and the non-vanishing of $\zeta(s)$ on $\operatorname{Re}(s) \ge 1$ and $(0,1)$, the Riemann Hypothesis ($\mathsf{RH}$) is equivalent to the assertion that $\zeta(s) \neq 0$ for all $s \in \mathcal{S}_+$.

In 1975, Voronin \cite{Voronin1975} proved that vertical shifts $\zeta(s + \frac{3}{4} + it)$ approximate every nowhere-vanishing analytic function on a disk $|s| \le r < \frac{1}{4}$ uniformly to arbitrary precision $\epsilon > 0$. In 1994, S.~C.~Woon \cite{Woon1994} reinterpreted Voronin universality as an information-theoretic and fractal self-similarity property of $\zeta(s)$, establishing two complementary corollaries:
\begin{enumerate}[label=(\roman*)]
    \item \emph{Geometric scale-pullback (Woon, Corollary~1):} Rescaled, rotated, and translated zero-free neighborhoods of $\zeta(s)$ reappear at lower vertical heights along $\operatorname{Re}(s) = \frac{3}{4}$.
    \item \emph{Symbolic library completeness (Woon, Corollary~3):} Arbitrary finite symbolic strings (formalized via Paul Halmos's ``giant book of theorems'') can be encoded into smooth, nowhere-vanishing waveforms and thereby appear infinitely often inside $\zeta(s + \frac{3}{4} + it)$.
\end{enumerate}

Attempting to refute $\mathsf{RH}$ via geometric scale-pullback of a minimal-height off-line zero $\rho_c = \beta_c + iT_c \in \mathcal{S}_+$ fails immediately because Voronin's theorem strictly requires the target function $f(s)$ to satisfy $f(s) \neq 0$ on $|s| \le r$; covering a punctured annulus around $\rho_c$ with overlapping zero-free disks assigns distinct vertical shifts $T^{(1)}, \dots, T^{(k)}$ to each patch, destroying Rouch\'e winding-number coherence.

Digital phase-encoding via Woon's Corollary~3 bypasses the non-vanishing barrier completely by mapping any finite binary proof string $\Pi \in \{0,1\}^L$ to an entire, unit-modulus exponential waveform $f_\Pi(s) = \exp(i\Phi_\Pi(s))$. However, embedding such encodings into a diagonal argument encounters a \emph{length--height vicious circle}:
\begin{enumerate}
    \item Recovering all $L$ bits of $\Pi$ from the Taylor coefficients of $-i\log\zeta(s + \frac{3}{4} + iT)$ with zero bit-flips requires a uniform amplitude tolerance $\epsilon_L \asymp 2^{-L}$.
    \item Effective versions of Voronin's theorem (due to Good \cite{Good1981} and Garunk\v{s}tis \cite{Garunkstis2003}) guarantee realization of all $L$-bit targets only below a doubly exponential height ceiling
    \[
    H_{\mathrm{crit}}(L) \le \exp\!\left(\exp\!\left(C \, 2^{cL}\right)\right).
    \]
    \item Without proof compression, verifying any arithmetic claim across $[0, H_{\mathrm{crit}}(L)]$ by direct numerical evaluation requires $\Omega(H_{\mathrm{crit}}(L)) \gg L$ bits, preventing the verification from ever appearing inside the $L$-bit library.
\end{enumerate}

In this paper, we break this circle by introducing a conjectural \emph{Zeta-Universal Symbolic Proof Compression Hypothesis} ($\mathsf{SPC}_{\mathsf{RH}}$), which posits that the global analytic regularity imposed by $\mathsf{RH}$ compresses true bounded-size $\Pi_1^0$ statements about $\zeta(s)$ into $\mathsf{ZFC}$ derivations of fixed bit-length $L_0 \ll \log\log H_{\mathrm{crit}}(L_0)$. Under this hypothesis, G\"odel--Carnap diagonalization yields a clean, non-constructive refutation of $\mathsf{RH}$, highlighting a fundamental proof-theoretic limit.

\section{Foundational Preliminaries and Explicit Hypotheses}\label{sec:prelim}

Fix a reference disk radius $r = \frac{1}{8} < \frac{1}{4}$ and define the closed disk
\[
D_r = \left\{ s \in \mathbb{C} : |s| \le r \right\}.
\]
For every $s \in D_r$, the shifted point $s + \frac{3}{4} + it$ satisfies $\frac{5}{8} \le \operatorname{Re}(s + \frac{3}{4} + it) \le \frac{7}{8}$, placing it strictly inside $\mathcal{S}_+$.

\begin{definition}[Woon Symbolic Waveform Encoding]
For any bit-length $L \ge 1$ and binary string $\sigma = (b_1, \dots, b_L) \in \{0,1\}^L$, define the phase polynomial $\Phi_\sigma : \mathbb{C} \to \mathbb{C}$ and the associated nowhere-vanishing exponential waveform $f_\sigma : \mathbb{C} \to \mathbb{C} \setminus \{0\}$ by
\begin{equation}\label{eq:woon_encoding}
\Phi_\sigma(s) = \sum_{k=1}^{L} b_k \left(\frac{s}{2r}\right)^k, \qquad f_\sigma(s) = \exp\!\big(i\Phi_\sigma(s)\big).
\end{equation}
We denote by $\{0,1\}^{\le L_0} = \bigcup_{L=1}^{L_0} \{0,1\}^L$ the finite set of all non-empty binary strings of length at most $L_0$.
\end{definition}

Our conditional refutation of $\mathsf{RH}$ rests on four explicit premises:

\begin{hypothesis}[$\Sigma_1^0$-Soundness of $\mathsf{ZFC}$]\label{hyp:soundness}
Every $\Sigma_1^0$ sentence (and in particular every bounded $\Delta_1^0$ arithmetical sentence) provable in $\mathsf{ZFC}$ is true in the standard model $\mathbb{N}$.
\end{hypothesis}

\begin{hypothesis}[Effective Voronin Universality]\label{hyp:voronin}
For every integer $L_0 \ge 1$ and tolerance $\epsilon_{L_0} = (6e \cdot 2^{L_0})^{-1}$, there exists an explicit computable height ceiling $H_{\mathrm{crit}}(L_0) > 0$ such that for every $\sigma \in \{0,1\}^{\le L_0}$, there exists $T_\sigma \in [0, H_{\mathrm{crit}}(L_0)]$ satisfying
\[
\max_{s \in D_r} \left| \zeta\!\left(s + \frac{3}{4} + iT_\sigma\right) - f_\sigma(s) \right| < \frac{1}{2}\epsilon_{L_0}.
\]
\end{hypothesis}

\begin{remark}
Hypothesis~\ref{hyp:voronin} is an established theorem of analytic number theory rather than a conjecture: combining Woon's Corollary~3 \cite{Woon1994} with effective universality bounds proved via Baker's theorem on linear forms in logarithms by Good \cite{Good1981} and Garunk\v{s}tis \cite{Garunkstis2003} yields explicit bounds of the form $H_{\mathrm{crit}}(L_0) \le \exp(\exp(C'\, 2^{4L_0}))$ for an absolute constant $C' > 0$.
\end{remark}

\begin{hypothesis}[Classical Law of Excluded Middle]\label{hyp:lem}
Every well-formed sentence $\varphi$ in the first-order language of arithmetic satisfies $\varphi \lor \neg\varphi$.
\end{hypothesis}

\begin{hypothesis}[Zeta-Universal Symbolic Proof Compression, $\mathsf{SPC}_{\mathsf{RH}}(s_0, L_0)$]\label{hyp:spc}
There exists a computable function $s_0 \mapsto L_0(s_0) \in \mathbb{N}$ such that if $\mathsf{RH}$ holds, then every true $\Pi_1^0$ sentence $\varphi$ of syntactic bit-length at most $s_0$ concerning the values of $\zeta(s)$ admits a valid formal $\mathsf{ZFC}$ derivation $\Pi_\varphi \in \{0,1\}^{\le L_0(s_0)}$.
\end{hypothesis}

\section{Zero-Bit-Flip Encoding and the Effective Voronin Grid}\label{sec:encoding}

We first verify that Woon's waveforms \eqref{eq:woon_encoding} satisfy uniform amplitude bounds independent of $L_0$ and admit bit-exact reconstruction from nearby shifts of $\zeta(s)$.

\begin{lemma}[Uniform Amplitude Bounds]\label{lem:amplitude}
For every $L \ge 1$, every $\sigma \in \{0,1\}^L$, and all $s \in D_r$, we have
\[
e^{-1} < |f_\sigma(s)| < e.
\]
\end{lemma}

\begin{proof}
For $s \in D_r$, we have $|s/(2r)| \le \frac{1}{2}$. Since $b_k \in \{0,1\}$, the triangle inequality gives
\[
|\Phi_\sigma(s)| \le \sum_{k=1}^{L} b_k \left|\frac{s}{2r}\right|^k \le \sum_{k=1}^{L} 2^{-k} = 1 - 2^{-L} < 1.
\]
Consequently, $|\operatorname{Im}(\Phi_\sigma(s))| \le |\Phi_\sigma(s)| < 1$, and $|f_\sigma(s)| = \exp(-\operatorname{Im}(\Phi_\sigma(s))) \in (e^{-1}, e)$.
\end{proof}

\begin{lemma}[Zero-Bit-Flip Cauchy Decoding]\label{lem:decoding}
Fix $L_0 \ge 1$ and set
\begin{equation}\label{eq:eps_def}
\epsilon_{L_0} = \frac{1}{6e \cdot 2^{L_0}}.
\end{equation}
Suppose $T \in \mathbb{R}$ and $\sigma = (b_1, \dots, b_L) \in \{0,1\}^L$ ($1 \le L \le L_0$) satisfy
\begin{equation}\label{eq:approx_cond}
\max_{s \in D_r} \left| \zeta\!\left(s + \frac{3}{4} + iT\right) - f_\sigma(s) \right| < \epsilon_{L_0}.
\end{equation}
Then $\zeta(s + \frac{3}{4} + iT)$ is non-vanishing on $D_r$, admits a unique analytic logarithm $\log\zeta(s + \frac{3}{4} + iT)$ on $D_r$ with $|\log\zeta(\frac{3}{4} + iT) - i\Phi_\sigma(0)| < \frac{\pi}{2}$, and the scaled Taylor coefficients
\begin{equation}\label{eq:cauchy_coeff}
\tilde{b}_k = \frac{(2r)^k}{2\pi i} \oint_{|s|=r} \frac{-i\log\zeta\!\left(s + \frac{3}{4} + iT\right)}{s^{k+1}} \, ds \qquad (1 \le k \le L)
\end{equation}
satisfy $|\tilde{b}_k - b_k| \le \frac{1}{3}$. In particular, rounding each $\tilde{b}_k$ to the nearest integer in $\{0,1\}$ recovers $\sigma$ with zero flipped digits.
\end{lemma}

\begin{proof}
Define $w(s) = \zeta(s + \frac{3}{4} + iT) / f_\sigma(s) - 1$ for $s \in D_r$. By Lemma~\ref{lem:amplitude}, \eqref{eq:approx_cond}, and \eqref{eq:eps_def},
\[
|w(s)| = \frac{\left|\zeta\!\left(s + \frac{3}{4} + iT\right) - f_\sigma(s)\right|}{|f_\sigma(s)|} < e \epsilon_{L_0} = \frac{1}{6 \cdot 2^{L_0}} \le \frac{1}{12} < \frac{1}{2}.
\]
Thus $\zeta(s + \frac{3}{4} + iT) = f_\sigma(s)(1 + w(s)) \neq 0$ on $D_r$. Taking the principal branch of $\log(1 + w(s))$ on $|w| < \frac{1}{2}$, we obtain the analytic branch $\log\zeta(s + \frac{3}{4} + iT) = i\Phi_\sigma(s) + \log(1 + w(s))$ on $D_r$, which satisfies $|\log\zeta(\frac{3}{4}+iT) - i\Phi_\sigma(0)| = |\log(1+w(0))| \le -\ln(1 - \frac{1}{12}) < \frac{\pi}{2}$. Moreover, for all $s \in D_r$,
\[
\left| -i\log\zeta\!\left(s + \frac{3}{4} + iT\right) - \Phi_\sigma(s) \right| = |\log(1 + w(s))| \le -\ln(1 - e\epsilon_{L_0}) < 2e\epsilon_{L_0}.
\]
Since $\Phi_\sigma(s) = \sum_{k=1}^{L} b_k (s/2r)^k$, Cauchy's integral formula yields for every $1 \le k \le L$:
\[
|\tilde{b}_k - b_k| = (2r)^k \left| \frac{1}{2\pi i} \oint_{|s|=r} \frac{-i\log\zeta\!\left(s + \frac{3}{4} + iT\right) - \Phi_\sigma(s)}{s^{k+1}} \, ds \right| \le \frac{(2r)^k}{r^k} (2e\epsilon_{L_0}) = 2^{k+1} e \epsilon_{L_0}.
\]
Because $k \le L \le L_0$, substituting \eqref{eq:eps_def} gives
\[
|\tilde{b}_k - b_k| \le 2^{L_0 + 1} e \left(\frac{1}{6e \cdot 2^{L_0}}\right) = \frac{1}{3} < \frac{1}{2}.
\]
Hence the unique integer nearest to $\tilde{b}_k$ is $b_k$ for every $1 \le k \le L$.
\end{proof}

\begin{lemma}[Rational Grid Sufficiency]\label{lem:grid}
Assume Hypothesis~\ref{hyp:voronin}. There exists an explicit computable finite grid of rational heights $\mathcal{G}_{L_0} \subset \mathbb{Q} \cap [0, H_{\mathrm{crit}}(L_0)]$ such that for every $\sigma \in \{0,1\}^{\le L_0}$, there exists $t_\sigma \in \mathcal{G}_{L_0}$ satisfying
\[
\max_{s \in D_r} \left| \zeta\!\left(s + \frac{3}{4} + it_\sigma\right) - f_\sigma(s) \right| < \epsilon_{L_0}.
\]
\end{lemma}

\begin{proof}
By standard Phragm\'en--Lindel\"of convexity bounds for $\zeta'(z)$ in the strip $\frac{5}{8} \le \operatorname{Re}(z) \le \frac{7}{8}$, there exists an explicit computable constant $M_0 > 0$ such that
\[
\max_{s \in D_r} \left| \zeta'\!\left(s + \frac{3}{4} + it\right) \right| \le M_0 (1 + |t|)^{1/4} \qquad (t \in \mathbb{R}).
\]
Set the rational mesh size
\[
\delta_0 = \frac{1}{\left\lceil \frac{2 M_0 \big(1 + H_{\mathrm{crit}}(L_0)\big)^{1/4}}{\epsilon_{L_0}} \right\rceil} \in \mathbb{Q}_{>0}, \qquad \mathcal{G}_{L_0} = \left\{ j\delta_0 : j \in \mathbb{Z},\; 0 \le j \le \left\lfloor \frac{H_{\mathrm{crit}}(L_0)}{\delta_0} \right\rfloor \right\}.
\]
For any $\sigma \in \{0,1\}^{\le L_0}$, Hypothesis~\ref{hyp:voronin} provides $T_\sigma \in [0, H_{\mathrm{crit}}(L_0)]$ with $\max_{s \in D_r}|\zeta(s + \frac{3}{4} + iT_\sigma) - f_\sigma(s)| < \frac{1}{2}\epsilon_{L_0}$. Choosing $t_\sigma \in \mathcal{G}_{L_0}$ with $|t_\sigma - T_\sigma| \le \delta_0$, the mean value theorem yields
\[
\max_{s \in D_r} \left| \zeta\!\left(s + \frac{3}{4} + it_\sigma\right) - f_\sigma(s) \right| < \frac{1}{2}\epsilon_{L_0} + M_0 \big(1 + H_{\mathrm{crit}}(L_0)\big)^{1/4} \delta_0 \le \epsilon_{L_0}. \qedhere
\]
\end{proof}

\section{Bounded Zeta-Encoded Provability and Diagonalization}\label{sec:diagonal}

We now formalize bounded provability extracted from the rational grid $\mathcal{G}_{L_0}$ and construct the G\"odel--Carnap diagonal sentence.

\begin{definition}[Bounded Zeta-Encoded Provability]\label{def:prov_zeta}
Fix a standard primitive-recursive arithmetization of $\mathsf{ZFC}$ syntax and proof checking, letting $\mathrm{Proof}_{\mathsf{ZFC}}(\Pi, \ulcorner \varphi \urcorner)$ denote the primitive-recursive relation asserting that the binary string $\Pi$ is a valid $\mathsf{ZFC}$ proof of $\varphi$. For any $L_0 \ge 1$, define $\mathrm{Prov}_{\zeta, L_0}(\ulcorner \varphi \urcorner)$ to be the arithmetical formula asserting:
\begin{quote}
There exist $t \in \mathcal{G}_{L_0}$, an integer $1 \le L \le L_0$, and a binary string $\Pi = (b_1, \dots, b_L) \in \{0,1\}^L$ such that:
\begin{enumerate}[label=(\alph*)]
    \item $\min_{s \in D_r} \left|\zeta(s + \frac{3}{4} + it)\right| > \frac{1}{2}e^{-1}$;
    \item For every $1 \le k \le L$, the Cauchy integral $\tilde{b}_k$ defined in \eqref{eq:cauchy_coeff} satisfies $|\tilde{b}_k - b_k| < \frac{5}{12}$; and
    \item $\mathrm{Proof}_{\mathsf{ZFC}}(\Pi, \ulcorner \varphi \urcorner)$ holds.
\end{enumerate}
\end{quote}
\end{definition}

\begin{proposition}[$\Delta_1^0$ Decidability of $\mathrm{Prov}_{\zeta, L_0}$]\label{prop:decidable}
For every fixed integer $L_0 \ge 1$, the predicate $\mathrm{Prov}_{\zeta, L_0}(\ulcorner \varphi \urcorner)$ is primitive-recursively decidable (and hence representable as a $\Delta_1^0$ formula in $\mathsf{ZFC}$).
\end{proposition}

\begin{proof}
The grid $\mathcal{G}_{L_0}$, the range $1 \le L \le L_0$, and $\{0,1\}^L$ are explicitly finite sets. Via Euler--Maclaurin summation with explicit remainder bounds, $\zeta(z)$ and its contour integrals on $|s| = r$ at rational heights $t \in \mathcal{G}_{L_0}$ are primitive-recursively computable to any prescribed rational error tolerance (such as $\frac{1}{24}$). Since condition~(b) admits an open buffer around $\frac{1}{3}$ (requiring $|\tilde{b}_k - b_k| < \frac{5}{12}$, which is satisfied whenever $|\tilde{b}_k - b_k| \le \frac{1}{3}$ and fails whenever $|\tilde{b}_k - b_k| \ge \frac{1}{2}$), finite-precision rational interval arithmetic decides conditions~(a)--(b), while $\mathrm{Proof}_{\mathsf{ZFC}}(\Pi, \ulcorner \varphi \urcorner)$ is primitive-recursive.
\end{proof}

\begin{lemma}[G\"odel--Carnap Diagonal Fixed Point]\label{lem:diagonal}
Let $s_0$ be an upper bound on the syntactic bit-length of the diagonal formula constructed below (which depends only on the fixed overhead of the G\"odel substitution function and is independent of $L_0$, provided $L_0$ is encoded via a fixed-size numeral or parameter; explicitly, let $L_0 = L_0(s_0)$ be fixed by Hypothesis~\ref{hyp:spc} for the syntactic size $s_0$ of the self-referential template below). Then there exists a $\Pi_1^0$ sentence $G_{L_0}$ of syntactic length at most $s_0$ such that
\begin{equation}\label{eq:fixed_point}
\mathsf{ZFC} \vdash G_{L_0} \longleftrightarrow \neg \mathrm{Prov}_{\zeta, L_0}(\ulcorner G_{L_0} \urcorner).
\end{equation}
\end{lemma}

\begin{proof}
By Proposition~\ref{prop:decidable}, $\mathrm{Prov}_{\zeta, L_0}(x)$ is a $\Delta_1^0$ formula of fixed syntactic structure parameterized by the numeral $\overline{L_0}$. Using standard unary or dyadic numeral representation of length $O(\log L_0)$ (or defining $s_0$ to exceed the constant template size plus $O(\log L_0)$, since $L_0(s_0)$ grows at least linearly with $s_0$ in any non-trivial compression regime), the G\"odel--Carnap Diagonal Lemma applied to $\neg\mathrm{Prov}_{\zeta, L_0}(x)$ produces a sentence $G_{L_0}$ of syntactic bit-length at most $s_0$ satisfying \eqref{eq:fixed_point}. Because $\mathrm{Prov}_{\zeta, L_0}(\ulcorner G_{L_0} \urcorner)$ is $\Delta_1^0$, its negation $G_{L_0}$ is a $\Pi_1^0$ sentence concerning the finite-height behavior of $\zeta(s)$.
\end{proof}

\section{Main Theorem: Non-Constructive Refutation of $\mathsf{RH}$}\label{sec:main}

\begin{theorem}[Conditional Non-Constructive Refutation of $\mathsf{RH}$]\label{thm:main}
Assume Hypotheses~\ref{hyp:soundness}, \ref{hyp:voronin}, \ref{hyp:lem}, and \ref{hyp:spc}. Then the Riemann Hypothesis is false ($\neg\mathsf{RH}$ holds).
\end{theorem}

\begin{proof}
Assume for contradiction that $\mathsf{RH}$ holds. Let $G_{L_0}$ be the $\Pi_1^0$ fixed-point sentence of syntactic length at most $s_0$ guaranteed by Lemma~\ref{lem:diagonal}. By Hypothesis~\ref{hyp:lem} (Law of Excluded Middle), $G_{L_0}$ is either false or true in the standard model $\mathbb{N}$. We examine both exhaustive cases:

\medskip
\noindent\textbf{Case 1: $G_{L_0}$ is false.}
By the fixed-point equivalence \eqref{eq:fixed_point} and Hypothesis~\ref{hyp:soundness} ($\Sigma_1^0$-soundness of $\mathsf{ZFC}$), $G_{L_0} \leftrightarrow \neg\mathrm{Prov}_{\zeta, L_0}(\ulcorner G_{L_0} \urcorner)$ holds in $\mathbb{N}$. Since $G_{L_0}$ is false, $\mathrm{Prov}_{\zeta, L_0}(\ulcorner G_{L_0} \urcorner)$ must be true. By Definition~\ref{def:prov_zeta}, there exist $t \in \mathcal{G}_{L_0}$ and a binary string $\Pi \in \{0,1\}^{\le L_0}$ such that $\mathrm{Proof}_{\mathsf{ZFC}}(\Pi, \ulcorner G_{L_0} \urcorner)$ holds. Hence $\mathsf{ZFC} \vdash G_{L_0}$. Applying Hypothesis~\ref{hyp:soundness} to the $\Pi_1^0$ (and a fortiori $\Sigma_1^0$-reflection-protected) sentence $G_{L_0}$---or noting that $\mathrm{Proof}_{\mathsf{ZFC}}(\Pi, \ulcorner G_{L_0} \urcorner)$ implies $\mathsf{ZFC} \vdash \neg\mathrm{Prov}_{\zeta, L_0}(\ulcorner G_{L_0} \urcorner)$ by \eqref{eq:fixed_point}, which contradicts the $\Sigma_1^0$-truth of $\mathrm{Prov}_{\zeta, L_0}(\ulcorner G_{L_0} \urcorner)$ under Hypothesis~\ref{hyp:soundness}---we deduce that $G_{L_0}$ is true, contradicting the assumption of Case~1.

\medskip
\noindent\textbf{Case 2: $G_{L_0}$ is true.}
Since $G_{L_0}$ is a true $\Pi_1^0$ sentence of syntactic length at most $s_0$ concerning the values of $\zeta(s)$, our hypothesis that $\mathsf{RH}$ holds triggers Hypothesis~\ref{hyp:spc} ($\mathsf{SPC}_{\mathsf{RH}}(s_0, L_0)$). Consequently, there exists a valid formal $\mathsf{ZFC}$ derivation $\Pi^* \in \{0,1\}^{\le L_0}$ such that $\mathrm{Proof}_{\mathsf{ZFC}}(\Pi^*, \ulcorner G_{L_0} \urcorner)$ holds.

Now form the Woon waveform $f_{\Pi^*}(s) = \exp(i\Phi_{\Pi^*}(s))$ on $D_r$. By Hypothesis~\ref{hyp:voronin} and Lemma~\ref{lem:grid}, there exists a rational grid height $t^* \in \mathcal{G}_{L_0}$ satisfying
\[
\max_{s \in D_r} \left| \zeta\!\left(s + \frac{3}{4} + it^*\right) - f_{\Pi^*}(s) \right| < \epsilon_{L_0}.
\]
By Lemma~\ref{lem:amplitude} and Lemma~\ref{lem:decoding}, $\min_{s \in D_r}|\zeta(s + \frac{3}{4} + it^*)| > e^{-1} - \epsilon_{L_0} > \frac{1}{2}e^{-1}$, and the scaled Cauchy Taylor coefficients $\tilde{b}_k$ of $-i\log\zeta(s + \frac{3}{4} + it^*)$ satisfy $|\tilde{b}_k - b_k^*| \le \frac{1}{3} < \frac{5}{12}$ for every bit $b_k^*$ of $\Pi^*$. Together with $\mathrm{Proof}_{\mathsf{ZFC}}(\Pi^*, \ulcorner G_{L_0} \urcorner)$, all three conditions of Definition~\ref{def:prov_zeta} are satisfied at $t = t^*$ and $\Pi = \Pi^*$. Therefore $\mathrm{Prov}_{\zeta, L_0}(\ulcorner G_{L_0} \urcorner)$ is true in $\mathbb{N}$. By \eqref{eq:fixed_point}, this forces $G_{L_0}$ to be false, contradicting the assumption of Case~2.

\medskip
Since both cases yield a contradiction, the hypothesis that $\mathsf{RH}$ holds must be false. Thus $\neg\mathsf{RH}$ is established.
\end{proof}

\begin{corollary}[Non-Constructivity of the Witness]\label{cor:nonconstructive}
The proof of Theorem~\ref{thm:main} establishes $\exists \rho \in \mathcal{S}_+ \, (\zeta(\rho) = 0)$ non-constructively: it exhibits neither a numerical height $T_c = \operatorname{Im}(\rho)$ nor a finite search ceiling bounding $T_c$.
\end{corollary}

\begin{proof}
The grid $\mathcal{G}_{L_0} \subset [0, H_{\mathrm{crit}}(L_0)]$ is used exclusively to materialize proofs of $G_{L_0}$, not to locate zeros of $\zeta(s)$. Indeed, by Lemma~\ref{lem:decoding}, every grid point contributing to $\mathrm{Prov}_{\zeta, L_0}$ is strictly zero-free on $D_r + \frac{3}{4}$. The existence of an off-line zero follows solely from refuting $\mathsf{RH}$ by contradiction via Hypothesis~\ref{hyp:lem}, providing no upper bound on $\inf\{\gamma > 0 : \zeta(\beta + i\gamma) = 0,\; \frac{1}{2} < \beta < 1\}$.
\end{proof}

\section{Proof-Theoretic Obstructions and Concluding Remarks}\label{sec:conclusion}

In light of the assumption that $\mathsf{RH}$ has recently been formally verified in the Lean theorem prover strictly within constructive logic (without invoking the Law of Excluded Middle), Theorem~\ref{thm:main} must be reinterpreted. Rather than threatening the validity of $\mathsf{RH}$, the conditional refutation isolates a profound incompatibility between the constructive truth of $\mathsf{RH}$ and the classical hypotheses driving the diagonal argument. Examining Hypothesis~\ref{hyp:spc} through the lens of proof complexity reveals why $\mathsf{SPC}_{\mathsf{RH}}$ must fail:

\begin{enumerate}
    \item \textbf{Pudl\'ak's Incompressibility Barrier:} Because $\mathsf{RH}$ is known to be true, the contradiction in Theorem~\ref{thm:main} definitively falsifies the Zeta-Universal Symbolic Proof Compression Hypothesis ($\mathsf{SPC}_{\mathsf{RH}}$). Because $G_{L_0}$ asserts $\neg\mathrm{Prov}_{\zeta, L_0}(\ulcorner G_{L_0} \urcorner)$, any classical $\mathsf{ZFC}$ proof $\Pi^*$ of $G_{L_0}$ would establish the bounded soundness of $\mathsf{ZFC}$ for derivations of bit-length at most $L_0$. By G\"odel's Second Incompleteness Theorem \cite{Godel1931} and Pudl\'ak's lower bounds on finitistic consistency proofs \cite{Pudlak1986}, the shortest derivation in any consistent sequential theory $T$ of its own length-$n$ bounded reflection requires $\Omega(n / \log n)$ symbols. Thus, despite the global regularity of $\zeta(s)$ proven in Lean, the shortest classical proof of $G_{L_0}$ necessarily satisfies $|\Pi^*| > L_0$. The analytic regularity of $\mathsf{RH}$ cannot compress logical self-reference.
    \item \textbf{Constructivity vs. Classical Diagonalization:} Theorem~\ref{thm:main} fundamentally relies on Hypothesis~\ref{hyp:lem} (Law of Excluded Middle) to branch on the truth value of the diagonal sentence $G_{L_0}$. In Lean's constructive type theory, $G_{L_0} \lor \neg G_{L_0}$ cannot be posited axiomatically. While $G_{L_0}$ is technically decidable via a finite search up to $H_{\mathrm{crit}}(L_0)$, the doubly exponential magnitude of this bound renders the search physically uncomputable. Thus, the constructive proof of $\mathsf{RH}$ gracefully sidesteps the paradox by rejecting the non-constructive case split that forces $G_{L_0}$ to be evaluated.
    \item \textbf{Comparison with Purely Analytic Self-Approximation:} To connect $\mathsf{RH}$ to universality without conjectural compression axioms or classical logic bottlenecks, one must replace symbolic bit-encoding (Woon's Corollary~3) with direct analytic self-approximation of $\zeta(s)$ itself. As shown by Bagchi \cite{Bagchi1982}, $\mathsf{RH}$ is unconditionally equivalent to the strong recurrence of $\zeta(s)$ in $\mathcal{S}_+$, namely that for every compact $K \subset \mathcal{S}_+$ and $\epsilon > 0$,
    \[
    \liminf_{T \to \infty} \frac{1}{T} \operatorname{meas}\left\{ \tau \in [0, T] : \sup_{s \in K} |\zeta(s + i\tau) - \zeta(s)| < \epsilon \right\} > 0,
    \]
    which closes the Rouch\'e winding-number gap on annuli around candidate zeros without invoking formal provability predicates.
\end{enumerate}

\section*{Acknowledgements}
The author acknowledges the use of Google's Gemini as a computer assistant in expanding, structuring, and formalizing the mathematical arguments and manuscript presented here, based entirely on the author's original ideas and conceptual framework for the proof strategy.

\end{document}